\documentclass{iaus}
\usepackage{graphicx}

\title[A-type stars] 
{A--type stars: evolution, rotation and binarity}

\author[Noels, Montalb\'an and Maceroni]   
{Arlette Noels$^1$, Josefina Montalb\'{a}n$^1$ \and Carla Maceroni$^2$ %
 }

\affiliation{$^1$Institut d'Astrophysique et G\'{e}ophysique Universit\'{e} de Li\`{e}ge,
\break All\'{e}e du 6 A\^{out}, B-4000 Li\`{e}ge, Belgium \break email: first.last@ulg.ac.be\\[\affilskip]
$^2$INAF - Osservatorio Astronomico di Roma, \break via Frascati 33, I-00040 Monteporzio C. (RM) - Italy 
\break email: maceroni@coma.mporzio.astro.it}

\pubyear{2004}
\volume{224}  
\pagerange{119--126}
\date{?? and in revised form ??}
\jname{The A-Star Puzzle}
\editors{J.\,Zverko, W.W.\,Weiss, J.\,\v{Z}i\v{z}\v{n}ovsk\'{y}, \& S.J.\,Adelman, eds.}
\begin{document}

\maketitle

\begin{abstract}
We discuss the internal structure of stars in the mass range 1.5 to 4\,$M_\odot$
from the PMS to the subgiant phase with a particular emphasis on the convective 
core and the convective superficial layers. Different physical aspects are considered
such as overshooting, treatment of convection, microscopic diffusion and rotation.
Their influence on the internal structure and on the photospheric chemical abundances
is briefly described.
   
The role of binarity in determining the observed properties and as a tool to 
constrain the internal structure is also introduced and the current limits of  
theories of orbital evolution and of available binary data--sets are discussed.
\end{abstract}

\firstsection 
\section{Introduction}
The theoretical evolution of A stars is extremely simple to compute if one ignores 
complex phenomena such as gravitational settling, 
radiative forces, rotation, turbulent mixing, magnetic fields, binarity... We 
shall briefly introduce the infuence of different treatments of convection, 
diffusion and rotation after a 
presentation of a "conservative" situation 
in which the only mixings come from convection and overshooting. Binarity as a 
tool to constrain stellar models is then discussed. 

\section{A--type stars: evolution}\label{sec:evol}
As A stars are located on or near the main sequence, we shall start this discussion 
with the core hydrogen burning phase. 

\subsection{Core hydrogen burning}
\begin{figure}
\vspace*{-1.5cm}

\centerline{\includegraphics[width=.5\textwidth]{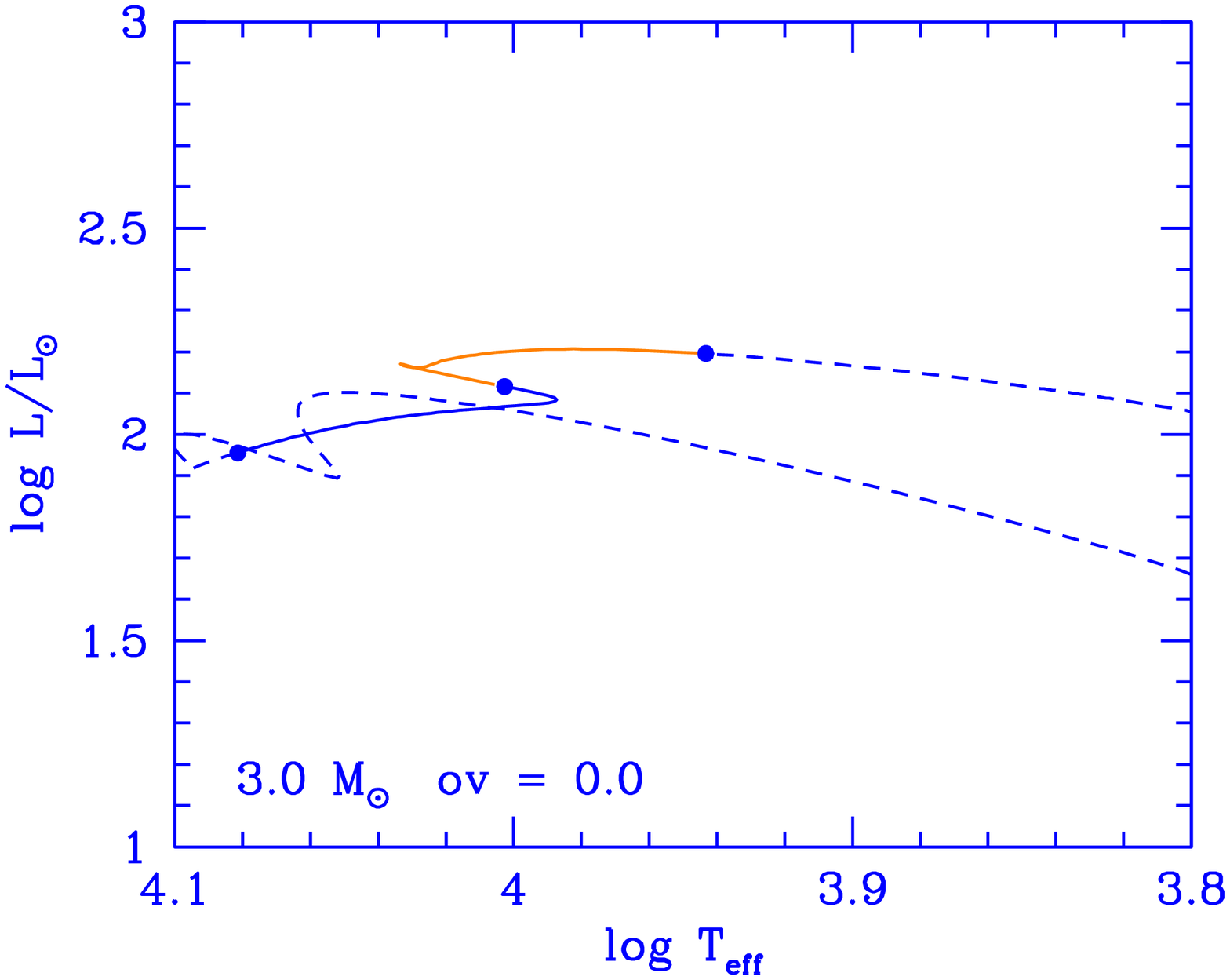}
 \includegraphics[width=.5\textwidth]{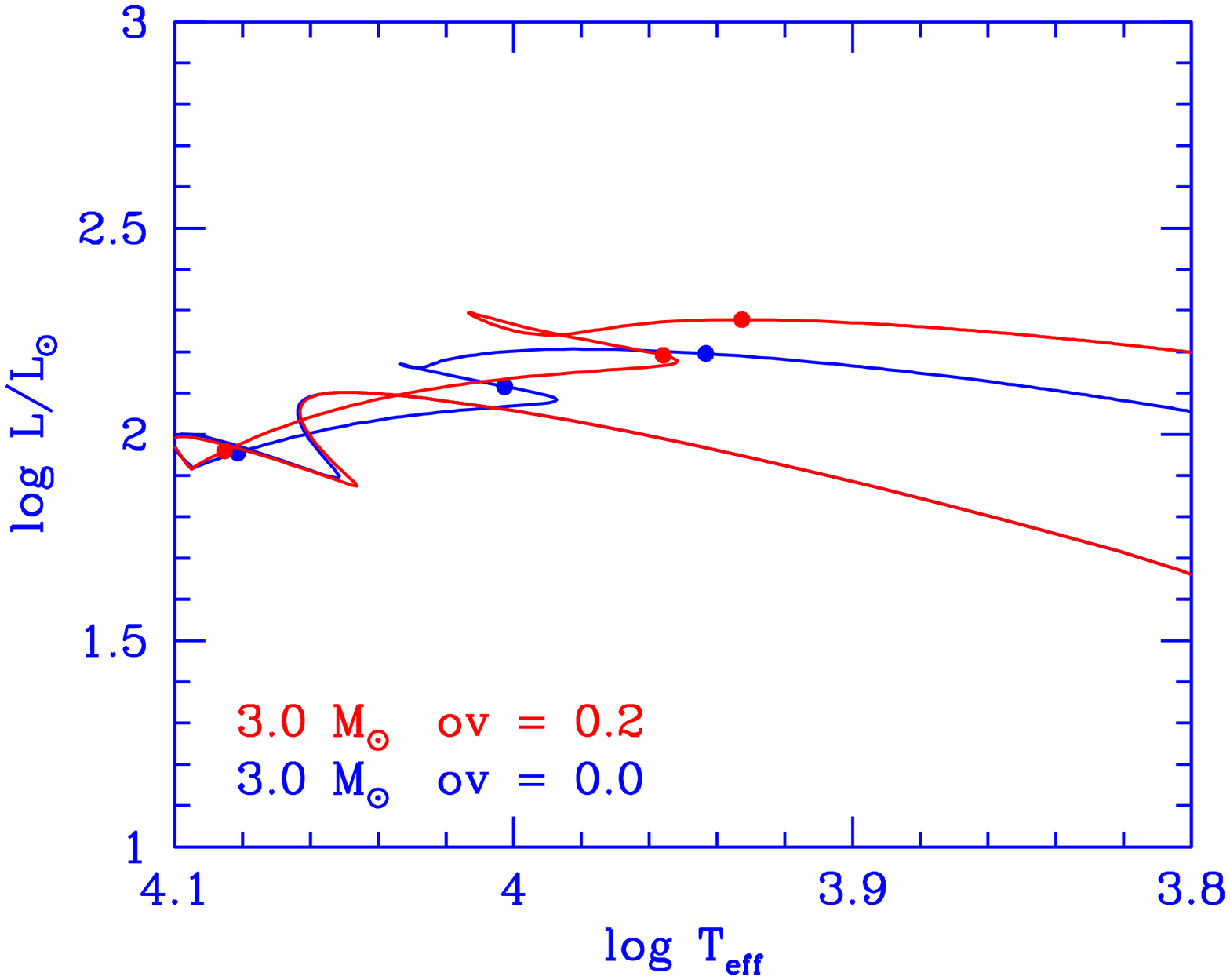}}
  \caption{Left panel (lp): Evolutionary track for a 3\,$M_\odot$ star computed 
  without overshooting. Right panel (rp): Same figure with an overshooting 
  parameter of 0.2 showing a wider MS track than in the left panel}
  \label{fig:evol}
\end{figure}

Figure~\ref{fig:evol} (lp) shows the evolutionary track of a 3\,$M_\odot$ with a 
solar chemical composition.

\begin{figure}
\vspace*{-1.5cm}

\centerline{\includegraphics[width=.5\textwidth]{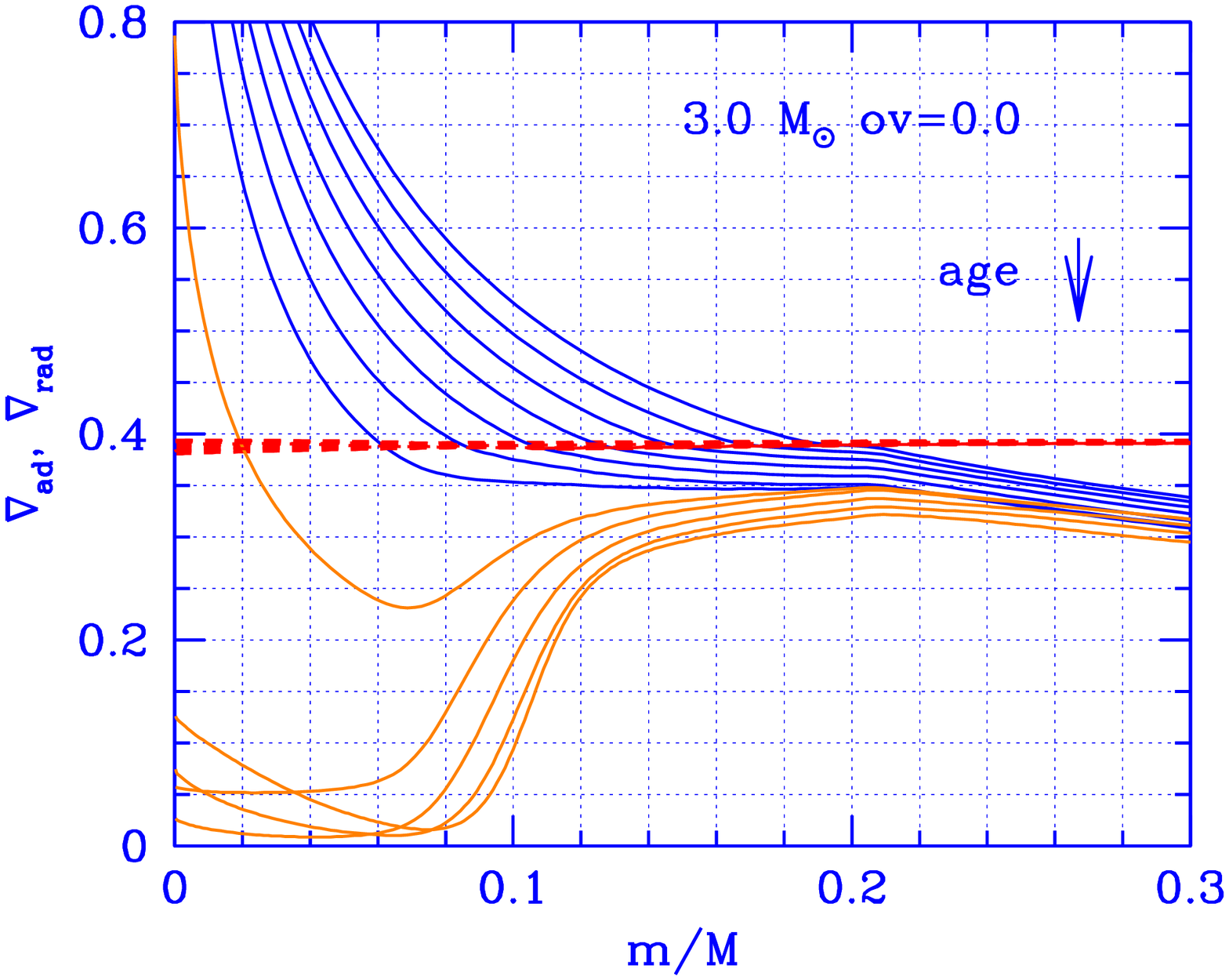}
\includegraphics[width=.5\textwidth]{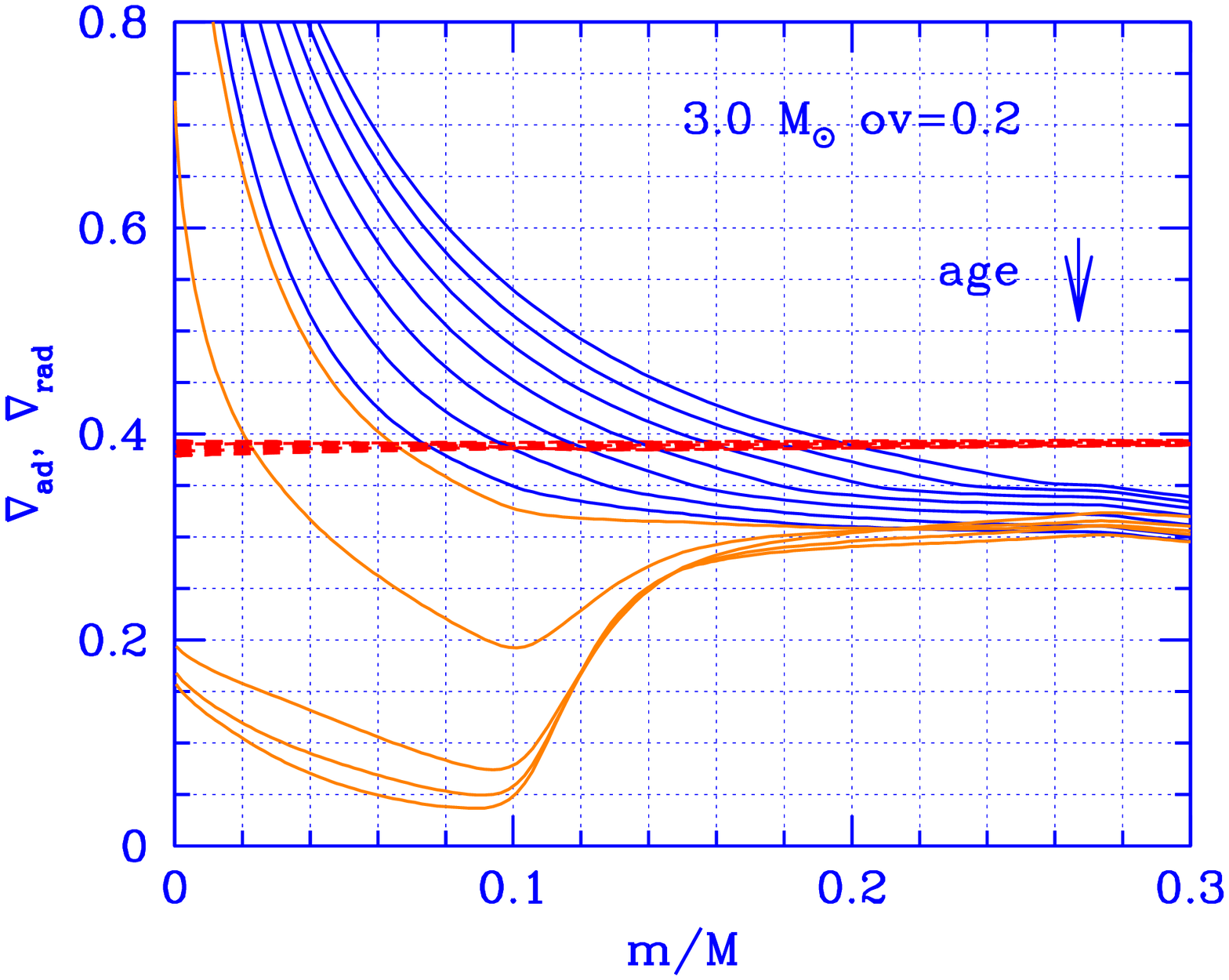}}
  \caption{Left panel (lp): Radiative gradient 
  for different models on the main sequence (black curves - first two dots starting 
  from the left in figure~\ref{fig:evol}) 
  and in the early H shell burning phase (gray curves - last two dots starting from 
  the left in figure~\ref{fig:evol}). 
  Right panel (rp): Same figure
  with an overshooting parameter of 0.2}\label{fig:rad}
\end{figure}

The evolution with time of the radiative 
gradient, given by
\begin{equation}
\bigtriangledown_{\rm rad} \sim \frac{L}{m} \; \kappa
\end{equation}
is shown in figure~\ref{fig:rad} (lp) in the inner 30 \% of the mass for selected 
models on the evolutionary path.

The adiabatic gradient, $\bigtriangledown_{\rm ad}$, is also drawn on this figure 
and its decrease towards the centre shows that the influence of the radiation 
pressure increases with time. 
Starting from the right of the figure, the radiative gradient increases when 
reaching the CNO burning central layers, due to the increasing value of $L/m$, 
and hydrogen burning thus takes place in a convective core. Inside this core, 
the hydrogen abundance, {\it X}, decreases. The opacity, $\kappa$, being proportional 
to (1+{\it X}), decreases as time goes on. This means that the convective core 
has its maximum extension in mass on the ZAMS and then slowly recedes as hydrogen 
is transformed into helium. 
The difference in {\it X} between 
the core and the non-burning regions thus increases with time without any discontinuity 
at the edge since the convective core recedes. A gradient of hydrogen builds up and 
a plateau in $\bigtriangledown_{\rm rad}$ appears (see figure~\ref{fig:rad} (lp)).  

\begin{figure}
\vspace*{-1.5cm}
  \centerline{\includegraphics[width=.5\textwidth]{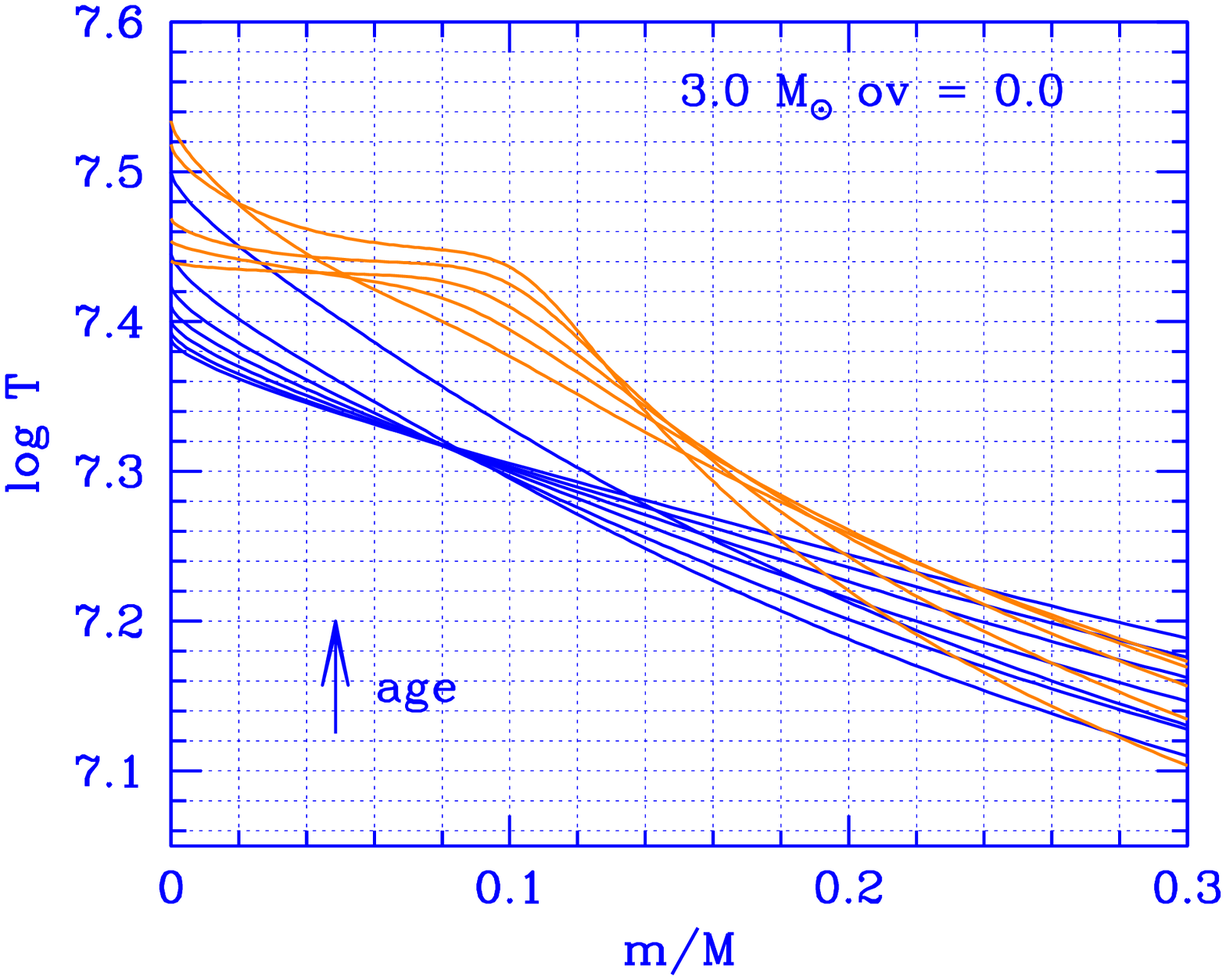}
\includegraphics[width=.5\textwidth]{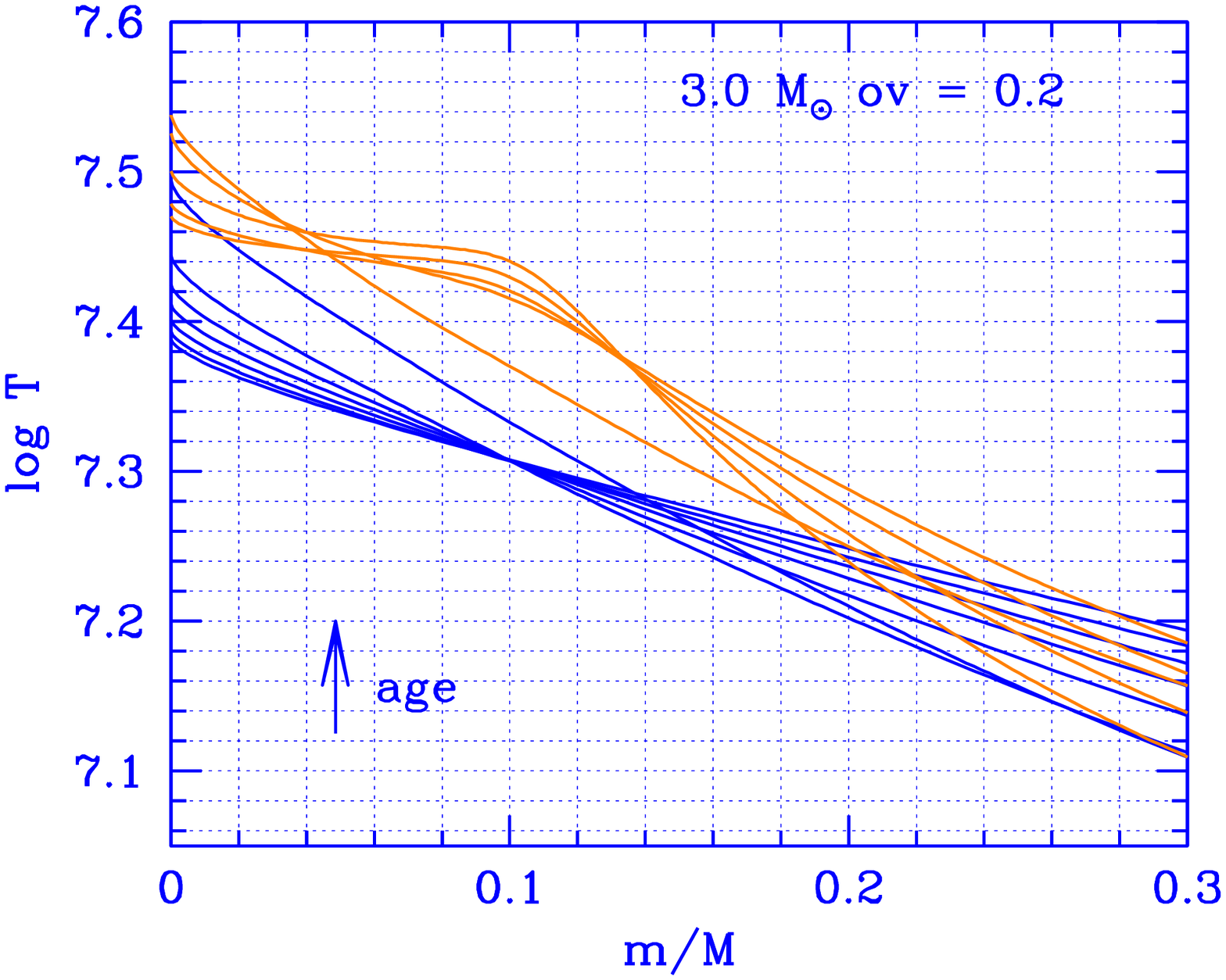}}
  \caption{Left panel (lp): Temperature versus mass 
  for different models on the main sequence (black curves - first two dots starting 
  from the left in figure~\ref{fig:evol}) 
  and in the early H shell burning phase (gray curves - last two dots starting from 
  the left in figure~\ref{fig:evol}). 
  Right panel (rp): Same figure
  with an overshooting parameter of 0.2}\label{fig:temp}
\end{figure}

The decrease of {\it X} leads to an increase in the temperature (Figure~\ref{fig:temp}
(lp)) which 
is limited to the central layers while the other layers become cooler and cooler.
\begin{figure}
\vspace*{-1.5cm}
  \centerline{\includegraphics[width=.5\textwidth]{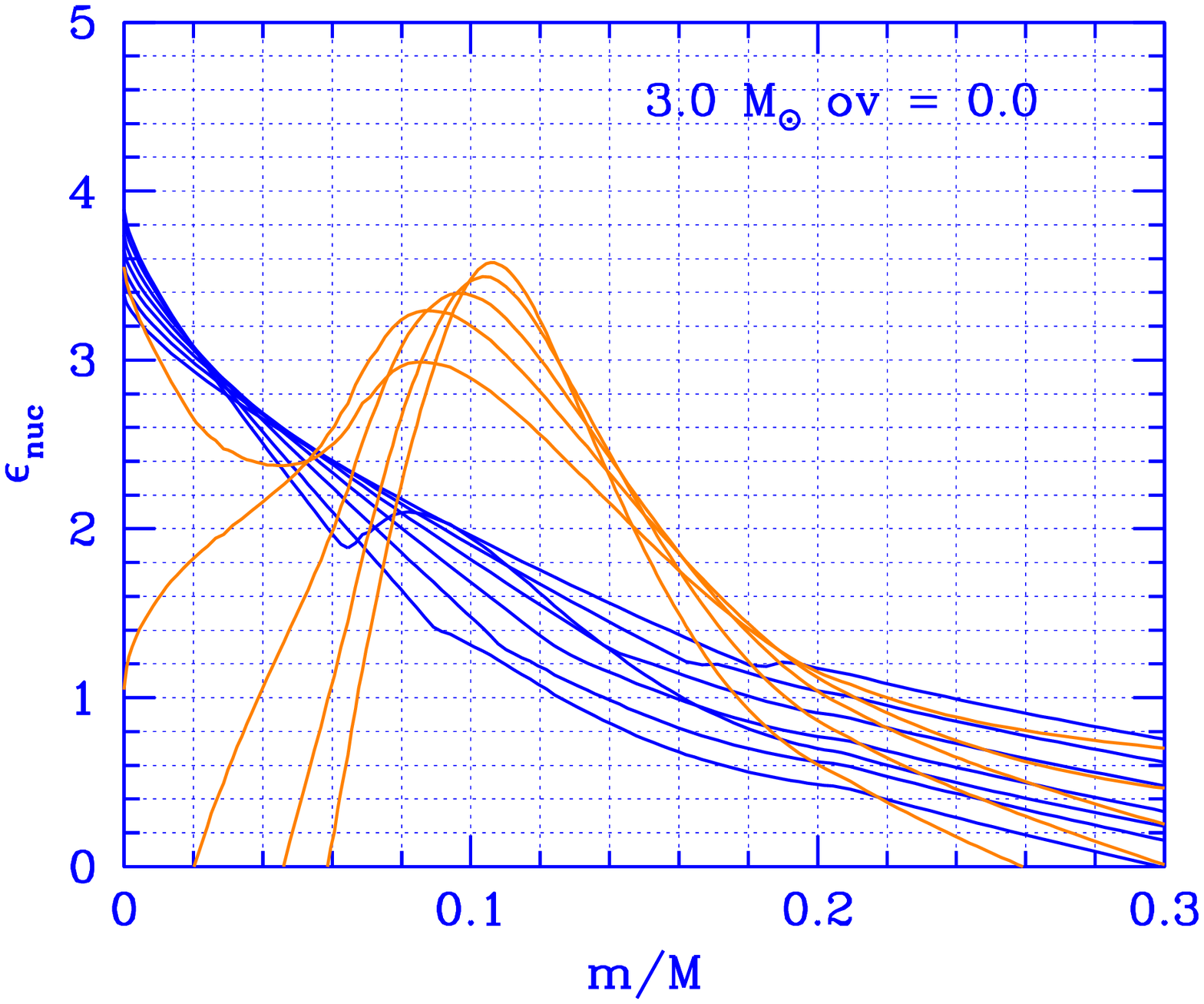}
\includegraphics[width=.5\textwidth]{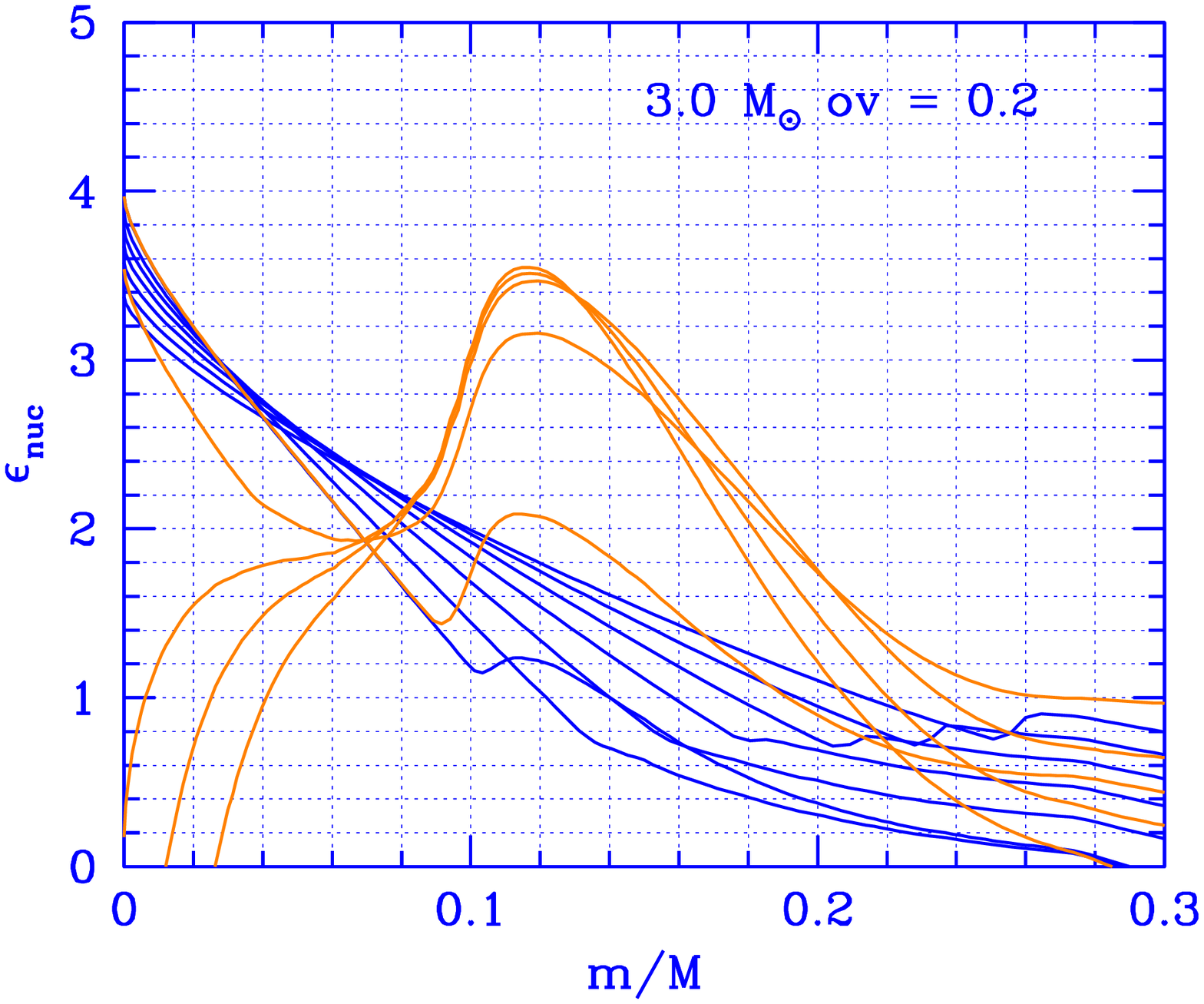}}
  \caption{Left panel (lp): Nuclear energy rate versus mass 
  for different models on the main sequence (black curves - first two dots starting 
  from the left in figure~\ref{fig:evol}) 
  and in the early H shell burning phase (gray curves - last two dots starting from 
  the left in figure~\ref{fig:evol}). 
  Right panel (rp): Same figure
  with an overshooting parameter of 0.2}\label{fig:eps}
\end{figure}
The central part of the star is contracting while the other layers are expanding. 
Once {\it X} comes close to zero in the core, the increase in temperature becomes more 
pronounced (see figure~\ref{fig:temp} (lp)). This is done by a global 
contraction (the so-called \it second gravitational contraction\rm).
\subsection{Hydrogen shell burning}
In the $\mu$-gradient zone, a hydrogen burning shell starts to develop. A secondary 
maximum is seen in $\bigtriangledown_{\rm rad}$ (Figure~\ref{fig:rad} (lp)) as well as 
in the nuclear enregy rate, $\epsilon$ (Figure~\ref{fig:eps} (lp)). Soon after, 
$\bigtriangledown_{\rm rad}$ becomes smaller than $\bigtriangledown_{\rm ad}$ and the convective 
core vanishes. As the energy 
production stops, an isothermal core appears. The models are now located between
the last two dots on the evolutionary track in figure~\ref{fig:evol} (lp). As hydrogen is 
burned in the shell, the 
isothermal core increases in mass. However, its mass cannot exceed the limiting 
mass of Sch\"{o}nberg-Chandrasekhar, given by
\begin{equation}
\left(\frac{m}{M}\right)_{\rm SC} \simeq 0.37 \; \frac{\mu_{\rm e}}{\mu_{\rm i}}
\end{equation} 
which is of the order of 0.08. When this value is reached, the core must contract in 
order to create a temperature gradient (see figure~\ref{fig:temp} (lp)) and this is 
accompanied by an 
expansion of the envelope where the temperature decreases. 
\subsection{Effect of overshooting}
In order to fit CM diagrams of open clusters as well as eclipsing binaries, 
overshooting seems to be needed with an extent increasing with stellar mass. 
We have computed a similar evolution with an overshooting parameter of 0.2. 
A larger hydrogen reservoir due to the increase of the mixed region 
evidently translates into a longer core hydrogen burning phase as well as in a longer
track in the HR diagram (figure~\ref{fig:evol} (rp)). However, after the turn-off, for 
a similar variation in 
effective temperature, the evolution with overshooting is much more rapid.
Figure~\ref{fig:rad} (rp) shows that $\bigtriangledown_{\rm rad}$ never reaches zero 
so there is no plateau in temperature (figure~\ref{fig:temp} (rp)). The reason 
lies in the fact that with the 
increase of the mixed region, the exhaustion of hydrogen takes place in 
a core whose mass is already higher than $\left(\frac{m}{M}\right)_{\rm SC}$. 
This results in a much quicker contraction
of the central regions. The star moves more rapidly and for a similar change in 
effective 
temperature, the shell, shown by a maximum in the distribution of $\epsilon$, 
is still at 
the same mass fraction with overshooting while
the shell has already moved significantly in mass without overshooting 
(compare lp with rp in figure~\ref{fig:eps}).
 
\it Although the $\mu$-gradient has a smaller slope with overshooting than without 
overshooting, the presence or absence of an isothermal core leads to a similar extent
of the helium core at a similar effective temperature. \rm
\subsection{Pre-main sequence evolution}
\begin{figure}
\centerline{\includegraphics[width=0.33\textwidth]{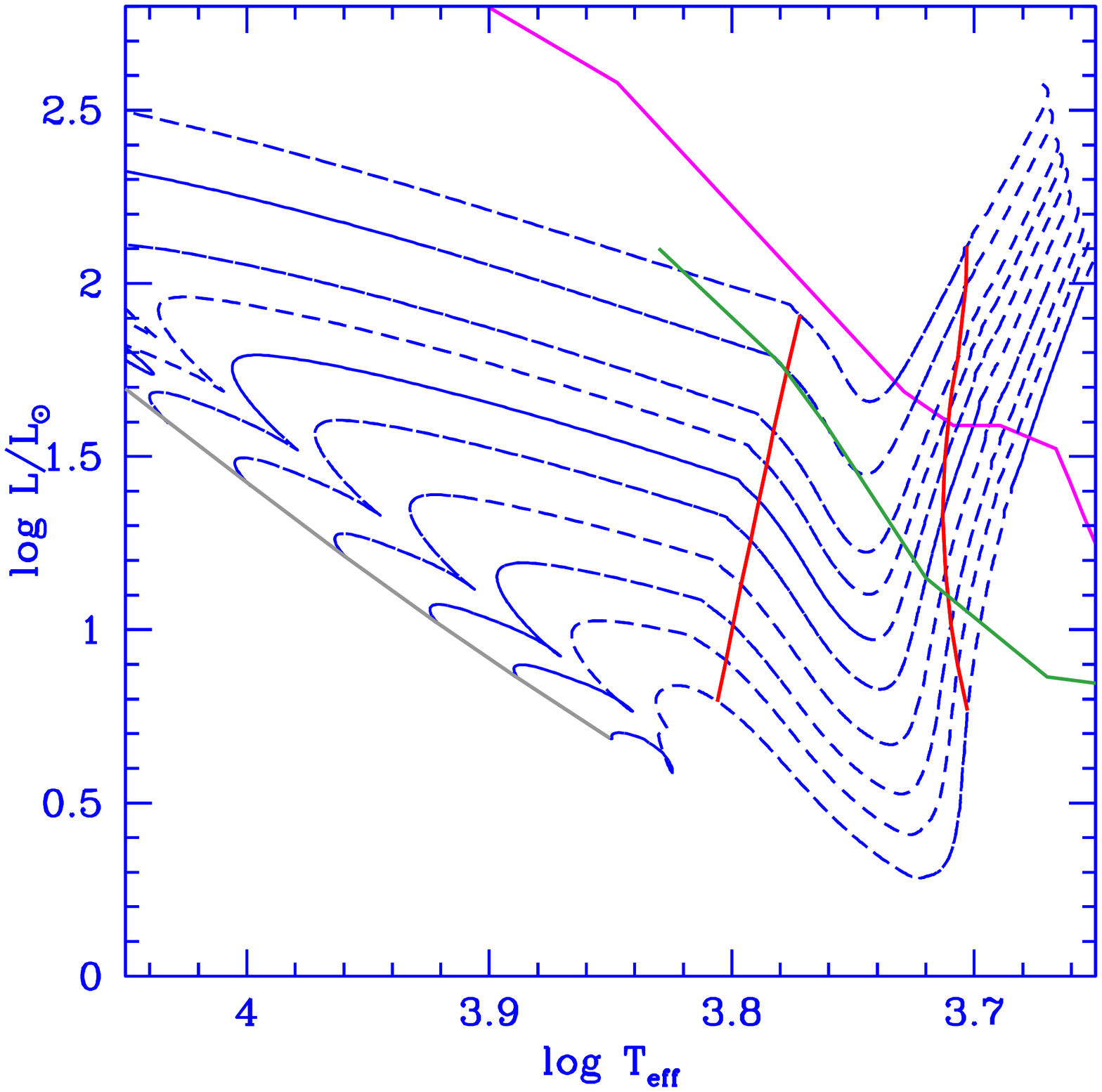}
\includegraphics[width=0.33\textwidth]{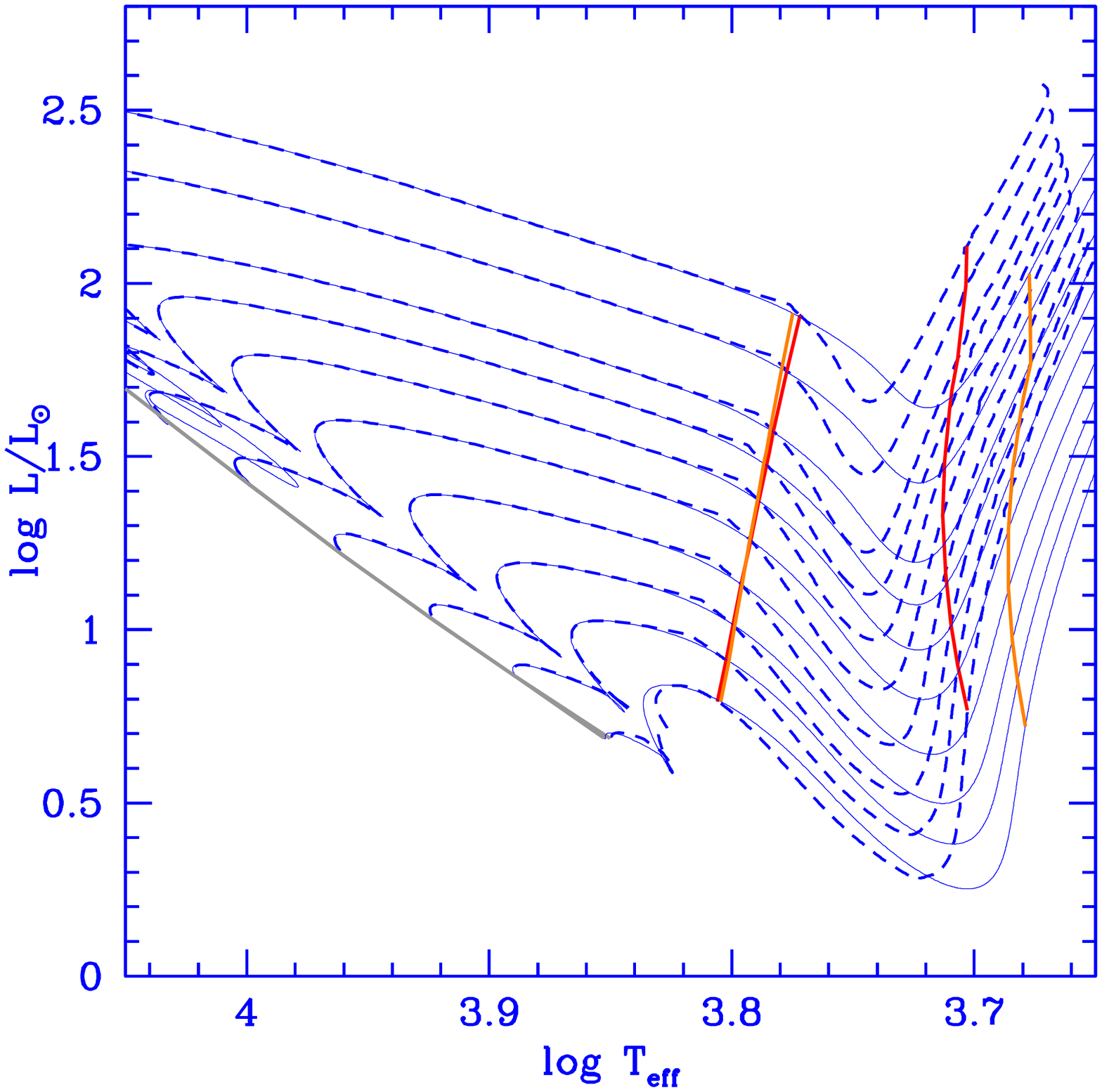}
\includegraphics[width=0.33\textwidth]{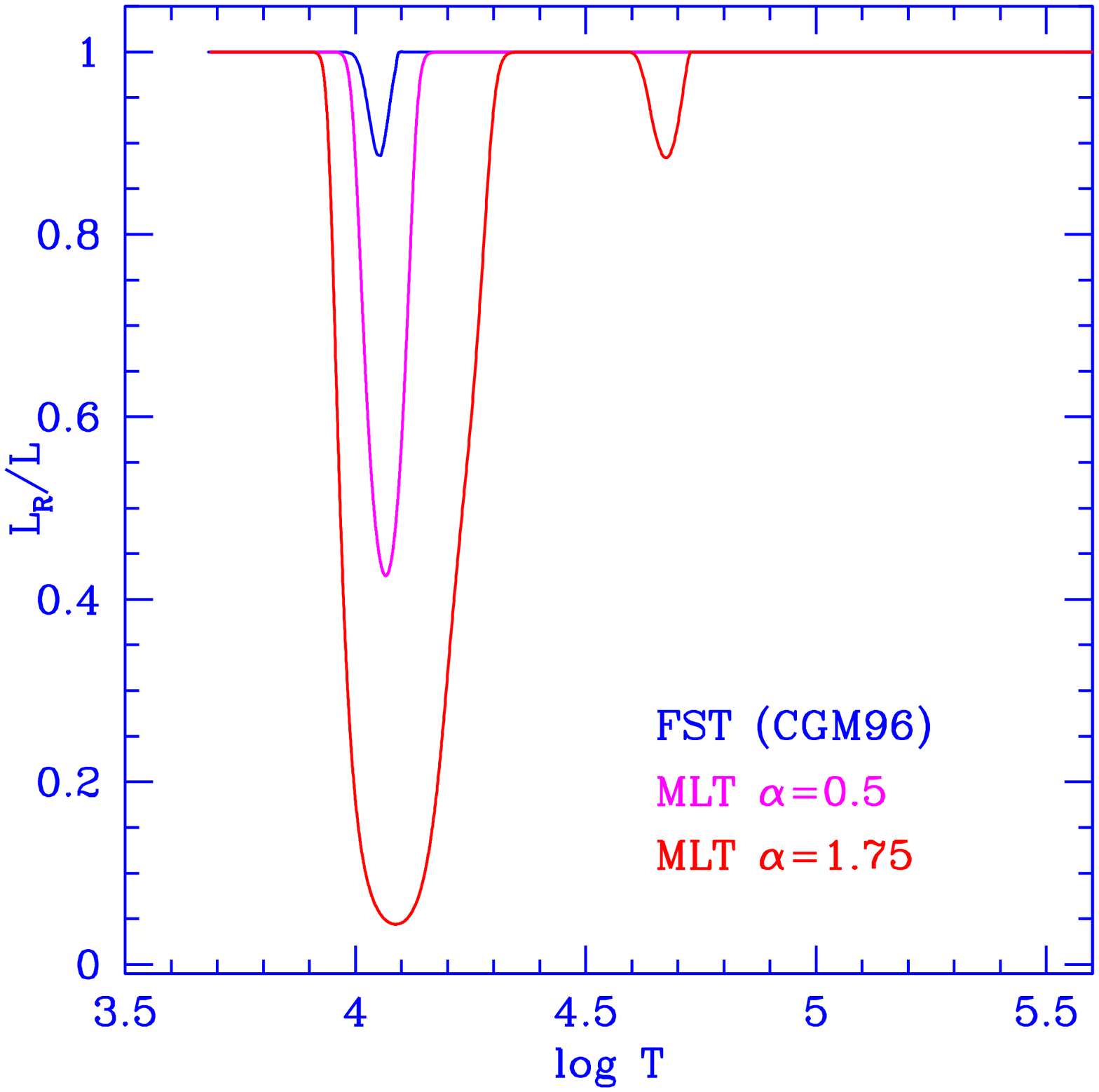}}
  \caption{Left panel (lp): PMS tracks (see text) with two birthlines (bottom one: 
  \cite{pal93}, top one: \cite{mae01}. Center panel (cp): 
  Evolutionary tracks computed
  with FST (dashed curves) and with MLT ($\alpha$ = 1.6) (faded curves). 
  Right panel (rp):
  Fraction of energy transported by radiation versus $\log T$ in the external 
  layers. (1) \it With FST~: \rm Only one (the smallest) minimum near $\log T = 4.1$ 
  (H {\sc i} and
  He {\sc i}) and no minimum near $\log T = 4.7$. (2) \it With MLT 
  ($\alpha$ = 0.5)~: \rm One 
  minimum (the intermediate one) near $\log T = 4.1$ and no minimum near $\log T = 4.7$. 
  (3) \it With MLT ($\alpha$ = 1.75)~: \rm One 
  (the largest) minimum near 
  $\log T = 4.1$ and a second small one near $\log T = 4.7$ (He {\sc ii})}
  \label{fig:pms}
\end{figure}

Let us now turn toward the pre-main sequence evolution (figure~\ref{fig:pms}). 
The two quasi-vertical lines on the Hayashi tracks and soon after 
show where the star becomes partly radiative and then completely radiative. The 
loops near the MS are the signatures of the CNO reactions evolving toward 
equilibrium. We have checked the effect of the new (smaller) cross-section of 
the slowest of the 
CNO reactions, i.e. $^{14}{\rm N}(p,\gamma)^{15}{\rm O}$ (\cite{luna04}). Although the approach 
toward equilibrium is slower, 
the difference in the main sequence lifetime is not significant. 

The points in the HR diagram where the stars become visible define the birthline.
Two birthlines are drawn in figure~\ref{fig:pms} (lp)).
The bottom one comes from \cite{pal93} and is constructed assuming a constant accretion rate of
$10^{-5}\,M_\odot$/yr. 
This value
was obtained by fitting the Herbig Ae and Be objects in the HR diagram. The top one 
comes from an analysis by \cite{mae01} where the accretion rate
\begin{equation}
\dot{M} \; = \; \frac{1}{3} \; \dot{M}_{\rm disk}
\end{equation}
has been derived in order to reproduce the observations in the HR diagram, in 
particular for the most massive stars. The time spent on the Hayashi 
track depends on the adopted birthline. This results in a different transfer of 
angular momentum from the star to the disk, which in turn can change the angular 
momentum at the beginning of the MS phase.     

In these fully or partly convective phases, the treatment of convection affects 
the location in the HR diagram, creating a significant difference in effective 
temperature. Figure~\ref{fig:pms} (cp) shows the location of the Hayashi tracks  
for models 
computed with the FST treatment of convection (dashed curves) and with the MLT treatment
($\alpha$ = 1.6) (faded curves). The difference in effective 
temperature
is of the order of 200\,K. As the convective envelope recedes, the tracks become 
undistinguishable. 

However, the extremely thin surface convection zone which remains on the main sequence
is affected (figure~\ref{fig:pms} (rp)).
The most superadiabatic the temperature gradient is, the less
efficient is the convection. For a 1.8\,$M_\odot$, it can make the 
convective He {\sc ii} ionization zone appear or disappear. This is important since the 
thickness of the mixed superficial layers is crucial in explaining the abundance 
anomalies.
\subsection{Gravitational settling}
\begin{figure}
  \includegraphics[height=3.5cm,width=4.3cm]{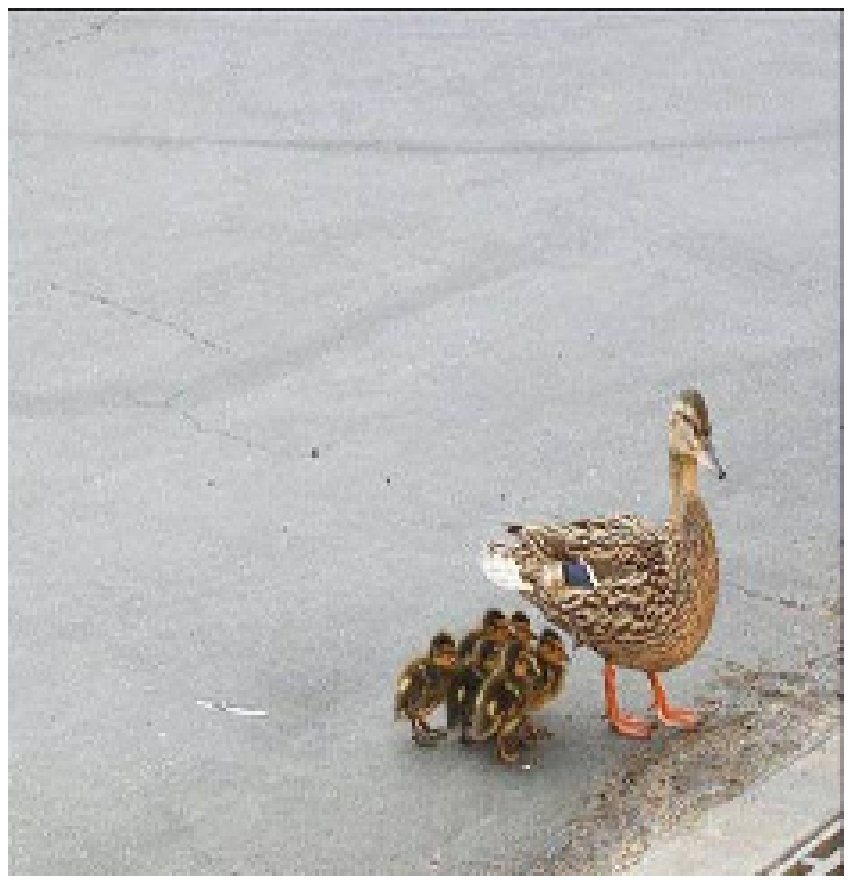}
  \includegraphics[height=3.5cm,width=4.3cm]{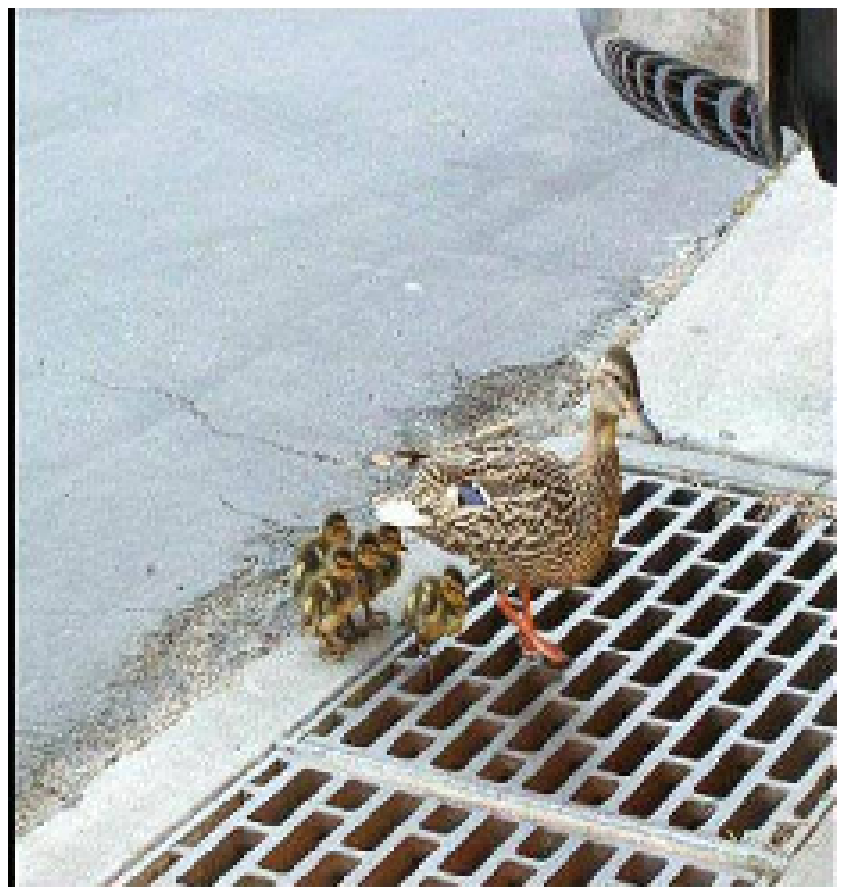}
  \includegraphics[height=3.5cm,width=4.3cm]{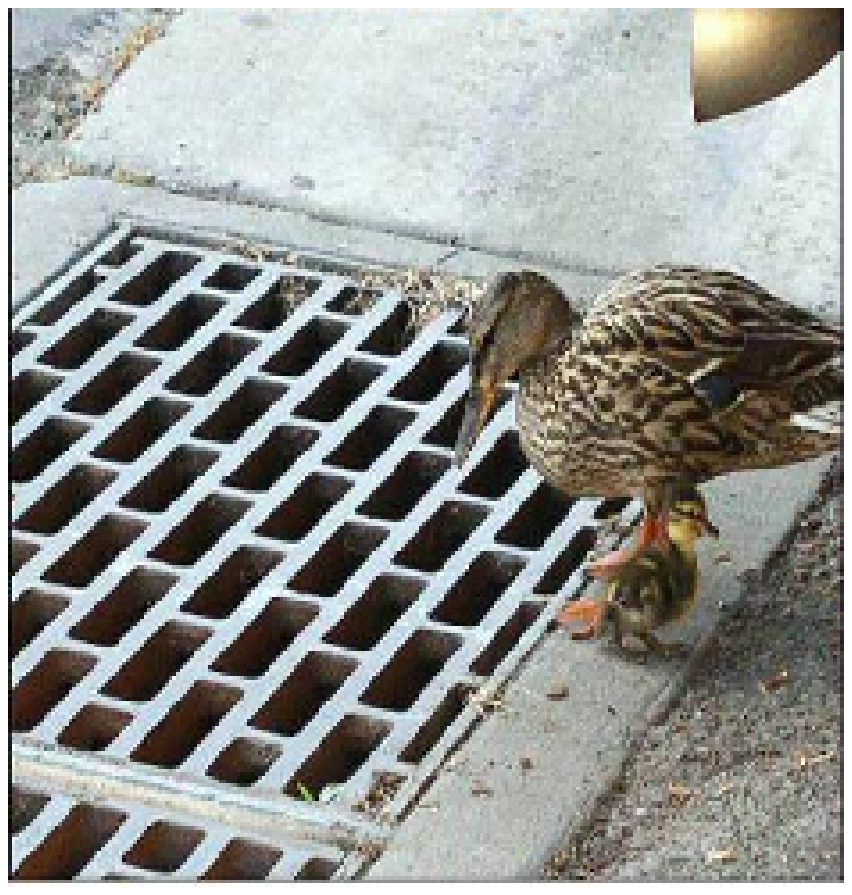}
  \caption{Illustration of diffusion. See text}\label{fig:can}
\end{figure}

With such thin convective envelopes, gravitational settling is very efficient. 
Figure~\ref{fig:can} illustrates this point.

In the left panel, a duck-star has all its chemical elements (chicken) showing at the 
surface. In the middle panel, the MS phase is symbolized by the crossing of a grid.  
If radiative forces are not taken into account together with gravitational settling, 
the duck-star ends its MS phase with only one element (chick) remaining at the surface. 

However, radiative forces can in turn be too efficient and another mechanism must 
enter the game. It is the turbulent mixing which will not only affect the surface 
abundances but will induce changes in the internal structure as well, especially in 
the thickness of the convective envelope (\cite{Richard01}). One striking example is the 
formation of a convective iron zone at a temperature of  200000\,K caused by an 
accumulation of iron-peak elements as a result of radiative forces. According to the 
mass of the star, this zone can merge with the H-He ionisation zone during the main 
sequence or remain detached, which affects the surface abundances of the iron-peak 
elements.




\section{A--type stars: rotation}\label{sec:rot}

 From their location in the HR diagram,  A-type stars are expected to rotate rapidly
and not to be affected by magnetic braking.  They can reach indeed  rotational velocities up to  300 km\,s$^{-1}$. 
Abt \& Morrel (1995, hereafter AM95) showed 
however, that the distribution of rotational velocities has a bimodal shape, with virtually 
all the Am and Ap stars  having equatorial rotational velocities ($v_{\rm rot}$) less than 120 km\,s$^{-1}$, 
and most of the  normal A0-F0 main-sequence stars  having $v_{\rm rot} > 120$~km\,s$^{-1}$.
They concluded (see also \cite{Abt00})  that  rotation alone can  explain the occurrence of abnormal
or normal main-sequence A stars, and that the apparent overlap between their $v_{\rm rot}$  distributions
is only due to our inability to distinguish marginal Am stars from normal ones, or to disentangle 
rotational and  evolutionary effects.

A large number of observational data indicates that during  MS evolution of A-type stars: 
 1)  mass loss is limited to $ \sim 2.10^{-10}$ $M_{\odot}{\rm yr}^{-1}$.
(Lanz \& Catala, 1992);
2) rotational velocity does not depend on age; and
3) no significant angular momentum loss by magnetic braking is observed 
(e.g. Wolff \& Simon , 1997; Hubrig, North \& Medici, 2000).                
Consequently,  the observed $v_{\rm rot}$ distribution must be determined by the angular momentum
evolution during the pre-main sequence (PMS) phase. In a study of the distribution of angular momentum
in Orion stars, \cite{Woff04} find  this hypothesis  consistent with a simplified model of PMS evolution in which
angular momentum is lost by interaction with the protostellar disk during  the convective phase, and 
conserved in the radiative one. Furthermore a core-envelope decoupling occurs during the convective-radiative
transition (see also \cite{step00} and \cite{step02} for Ap PMS angular momentum evolution).
The understanding of  angular momentum evolution during the PMS and the MS is fundamental 
to understand the Ap and the  Am phenomena.

\subsection{Modeling the evolution of a rotating star}

Rotation has several different effects on stellar evolution: change of the internal hydrostatic equilibrium, 
changes in the apparent effective temperature and luminosity (P\'erez Hern\'andez {\it et al.} 1999 and references therein); 
transport  of chemicals and of angular momentum by shears in differentially rotating stars, by meridional circulation 
an by horizontal turbulence (e.g., Zahn, 1974; Knobloch \& Spruit, 1982).

The effect of rotation on the stellar evolution has been treated with different approaches:  Endal \& Sofia (1981)
assigned to each  transport process (including  meridional circulation)  a diffusion coefficient, while 
 \cite{Z92a} and \cite{MZ98} made  the 
hypothesis  that differential rotation in the radiative zone of a non-magnetic star gives rise 
to anisotropic turbulence (much stronger in the horizontal direction
 than in the vertical  one due to the stratification), and  those result in  ``shellular" rotation. 
 In this model the effective diffusion coefficient 
for the chemicals is:

\begin{equation}
D_{\rm eff}=\frac{|r U(r)|^2}{30\,D_{\rm h}}, {\rm with}\,\,  D_{\rm h}=\frac{r}{C_{\rm h}}\left|\frac{1}{3\rho r}\frac{{\rm d}(\rho r^2 U)}{{\rm d}r} - \frac{U}{2}\frac{{\rm d ln} r^2 \Omega}{{\rm d ln} r}\right|.
\end{equation}

\noindent Where $U$ is the meridional circulation velocity, $C_{\rm h}$ is a free parameter
 related to the turbulent horizontal viscosity, and the other quantities have the habitual meaning. 
\cite{palacios03}  has applied the Zahn's modeling of rotation with the formulation 
by \cite{MZ98} (including also microscopic diffusion but not radiative accelerations) 
to   masses from 1.35 to 2.2 ${\rm M}_{\odot}$.
We  recall here two results of this application: 1) the rotation profile inside the star shows  differential 
rotation mainly close to the  convective core. 2) The models (computed without overshooting), show a wider
MS when rotation is included. However, while the lower mass models increase their MS lifetime
by a 20\%, the highest masses increase it  by only  10\% (see their Fig.~1). This trend is opposite to that  
observed in open clusters  and in binary systems, where the fitting of  observations requires
to increase the overshooting parameter with the mass (i.e. \cite{Andersen90}; Ribas, Jordi \& Gim\'enez, 2000).

Recently, \cite{Maeder03} and independently Richard \& Zahn (1999)  have updated the  value for the horizontal
 diffusion coefficient, $C_{\rm h}$ by a factor of $10^2$ with respect to the value used in \cite{MZ98}.
 Its implementation in  a stellar evolution code (Mathis, Palacios \& Zahn 2004) leads to: 
1) enhanced mixing, and  2)  significant changes  in the profile of chemical mixing with depth (see their Fig.2). 
To which degree will the new $D_{\rm eff}$ (with a non-zero value until the boundary of the convective core) 
 affect the MS width  and the surface chemical abundances?
 
The scheme proposed by AM95 matches qualitatively well  the predictions of the microscopic diffusion models
(e.g. Richer, Richard \& Michaud 2000), in the hypothesis that the extra-mixing required to decrease
the  microscopic diffusion is induced by rotation.  
Some problems are however left, for instance:  1)  the ``normal'' late B  and early A-type
 main-sequence stars show a true bimodal $v_{\rm rot}$ distribution (Royer, G\'omez \& Zorec, these proceedings); 2)
 the binary V392 Carinae  has a  $v\sin i$ of  27 km\,s$^{-1}$  and no peculiar abundance (Debernardi \& North 2001).
3) There is  no correlation between the strength of chemical peculiarities and $v_{\rm rot}$
 (\cite{EspNor03}).  Models including microscopic diffusion and
radiative accelerations show that the chemical abundances are very sensitive to the thickness of the 
mixed layer (e.g., Alecian 1996, \cite{HBH00}; \cite{Richard01}), so that we  should expect 
a signature of $v_{\rm rot}$ on the abundances, if  rotation is the responsible
of the extra-mixing below the convective envelope. 
A first approach to explicitly include a model of rotationally induced mixing and a complete treatment
of microscopic diffusion is in progress (Richard, Talon \& Michaud, these proceedings).

Finally, there are also other effects that should be taken into account in the study of the evolution of
a rotating A-type star: 
\par
 1. {\it Interaction convection-rotation.} Rotation can reduce the efficiency of convection and  
 is able to reduce the extension of the overshooting region (e.g. \cite{julien97}). Recent numerical simulations 
of the a rotating convective core  (Browning, Brun \& Toomre, 2004) show that rotation leads to a variation of 
convective overshooting and penetration, and induces internal waves, meridional circulation, and differential 
rotation at the core boundary. 

\par
2. {\it Magnetic field.} \cite{MM03} have shown that
the  Tayler-Spruit magnetic instability (Spruit 2002) could take place in the interior of stars with small magnetic
fields and differential rotation, induce a process of angular momentum transport much more efficient than 
that due to the meridional circulation and horizontal turbulence and lead to solid body rotation on a short timescale.
Furthermore, in Ap stars, which have suffered  strong braking during their PMS phase (St{\c e}pie{\' n}, 2000),
differential rotation together with magnetic field could lead to magneto-rotational instabilities. These  
could transport angular momentum from the interior to the surface (\cite{arlt03}, and Artl, these proceedings), 
and  result in solid body rotation of these stars, as expected from some observational data
(Hubrig, North \& Medici, 2000).

\section{A--type stars: binarity}\label{sec:bin}
Binarity plays a fundamental role in the origin  and definition of A-type 
stars chemical peculiarities.   
The pioneering work of  Abt (1961, 1965)  showed  that most (he even
suggested all) Am type stars are relatively short period binaries  and that the 
period  distribution of binaries with Am and with non-peculiar A-type components
are complementary. Normal A-type components are found in systems with 
period shorter than $\sim 2.5$ or longer than $\sim 100$  days, while Am stars form 
binaries with period in the range 2.5--100 days.  
  
  Abt's  statements have somewhat lost of strength with time,
and with the increasing size of studied  binary samples. Some
overlap between the period distribution of normal and Am binaries
was found by the same author (AM95) and the binary frequency among Am 
stars has steadily decreased with time (see \cite{AL85} and \cite{Noal98}, 
\cite{Deb01} for results based on CORAVEL surveys). The latter author 
finds a frequency as low as 57\% (but CORAVEL samples are certainly biased 
against fast rotation  and, because of the limited
time span of the survey, against longer period--eccentric orbit binaries).
The -- at any rate -- high frequency has a straightforward explanation: in close
binaries the spin--orbit synchronization by tidal mechanisms can efficiently
brake the stellar rotation to  values compatible with the Am phenomenon.
 
 The other relevant connection between binarity and  A-type star peculiarity 
is in the low frequency of spectroscopic binaries among magnetic Ap stars
(\cite{AS73}, \cite{Gal85}, \cite{Noal98}). The current explanation is that
the strong magnetic fields prevent close binary formation (however, \cite{Bud99}
suggested instead that it is binarity to affect magnetisms and not the other way
around). 

While the general outline of the binarity--peculiarity connection is well established,  
there are still shortcomings in the theory of tidal synchronization. Besides,  
possible drawbacks in the interpretation of the observations  are caused by the 
origin and composition of the observed samples (biases, selection effects).

 Two  competing theories were developed to explain the observed levels of
orbital circularization and spin--orbit synchronization in close binaries: the
tidal theory of \cite{Z92} and the hydrodynamical theory of \cite{TT92}.
The necessary ingredients of the first are tidal bulges and an efficient 
dissipation mechanisms; in absence of synchronism, dissipation causes a lag of
the bulge  and hence a torque, which tends to establish synchronization
(and orbit circularization). The dissipation mechanisms at work are different 
for late and early type stars, in the first case it is turbulent dissipation in
the convective envelope  retarding  the equilibrium tide, in the second it is
radiative damping acting on the dynamical tide (forced gravity waves are emitted from 
the lagging convective core and are damped in the outer layers).

Tassoul's theory is based, instead, on the idea that while the torque due to dissipation 
processes is negligible,  transient strong meridional currents are produced by the tidal 
action and transfer angular momentum between the stellar interior and an ``Ekman  layer" 
close to the surface.  As a consequence if the  rotation period  is shorter  than the 
orbital one, the star is braked.   

Both theories yield time-scales for synchronization ($t_{m s}$) and circularization 
($t_{m c}$). 
For early type stars, in Zahn's case:
\begin{equation}
\label{zahn}
\frac{1}{t_{m s}}\propto \left ( \frac{GM}{R^3} \right )^{\frac{1}{2}} q^2 
(1+q)^{\frac{5}{6}} \frac{MR^2}{I} E_2 \left ( \frac{R}{a} \right )^{\frac{17}{2}}\!\!; \;\;\;\;
\frac{1}{t_{m c}}\propto \left ( \frac{GM}{R^3} \right )^{\frac{1}{2}} q 
(1+q)^{\frac{11}{6}} E_2 \left ( \frac{R}{a} \right )^{\frac{21}{2}}
\end{equation} 
where $q$  and $a$ are the mass ratio and the semi-axis of the binary, $M$, $R$, $I$  the mass, the 
radius and the inertial moment of the braked star and $E_2$ a constant related to the size of its 
convective core. In Tassoul's theory:
\begin{equation}
\label{tassoul}
\frac{1}{t_{m s}}\propto 10^{\frac{N}{4}-\gamma} q (1+q)^{\frac{3}{8}} \!  \left (
\frac{ML^2}{R^9} \right )^{\frac{1}{8}} \!\!\! \left ( \frac{R}{a} \right )^{\!\frac{33}{8}} \!\!; \;\;
\frac{1}{t_{m c}}\propto 10^{\frac{N}{4}-\gamma} (1+q)^{\frac{11}{8}} \beta^2 
\left ( \frac{ML^2}{R^9} \right ) ^{\frac{1}{8}} \!\!\! \left ( \frac{R}{a} \right )^{\!\frac{49}{8}}
\end{equation} 
with $L$ being the star luminosity, $\beta$ its fractional gyration radius, $N$ the ratio
between eddy and radiative viscosity ($N=0$ for radiative envelopes). The other factor, $\gamma$, was 
introduced by \cite{CGC95}, to take somehow into account the fact that the orbital evolution is derived  
(as in Zahn's case) under the hypothesis of small deviations from synchronism and from circular orbit. 
The circularization time is  in general two/three orders of magnitude longer than that of 
synchronization, due to  the larger amount of angular momentum stored in the orbit.
It is evident by comparing Eqs.~\ref{zahn} with \ref{tassoul} that Tassoul's mechanism has a 
longer range  and a much higher efficiency, especially 
for early-type stars. 
The timescales are -- at any rate -- {\it only an indication} of the process speed, and cannot 
replace the full integration of the orbital evolution equations. 

   In principle the comparison between the expected and the 
observed degree of synchronization/circularization in binaries of known age and 
 accurate dimensions could discriminate between the two theories, and  early type
stars are the best choice for such a test. However, the detailed treatment of   
\cite{CGC95} and \cite{CC97} (with simultaneous computation of orbital and stellar
evolution and a comparison sample of observed  binaries with the best known parameters) 
showed that the results are still inconclusive, as both models can explain only a part of 
the observations. Zahn's mechanism is not efficient enough for early type stars 
(but the pre-MS phases were not considered in the abovementioned papers) while 
Tassoul's  one is too efficient, unless a high value of $\gamma$ is introduced 
(to shorten the time-scales by a factor $\sim 40$). The most promising scenario 
remains therefore that suggested in a study of circularization in binaries with 
A-type components by \cite{MM92}: a process composed by an important  PMS phase 
of orbital evolution  followed by a MS phase, which however still awaits  full modeling.
 
\begin{figure}
\centerline{\includegraphics[width=.5\textwidth]{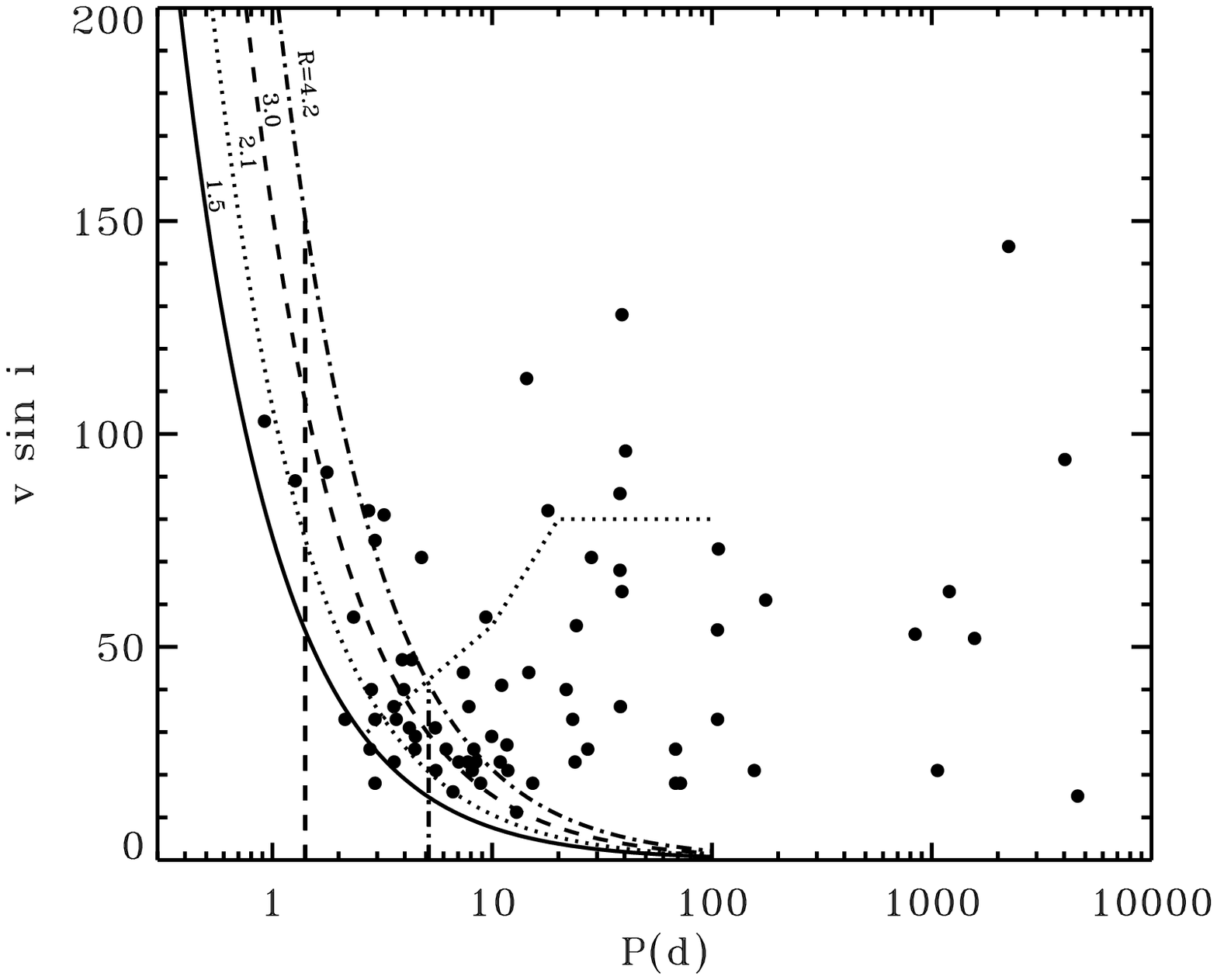} 
\includegraphics[width=.5\textwidth]{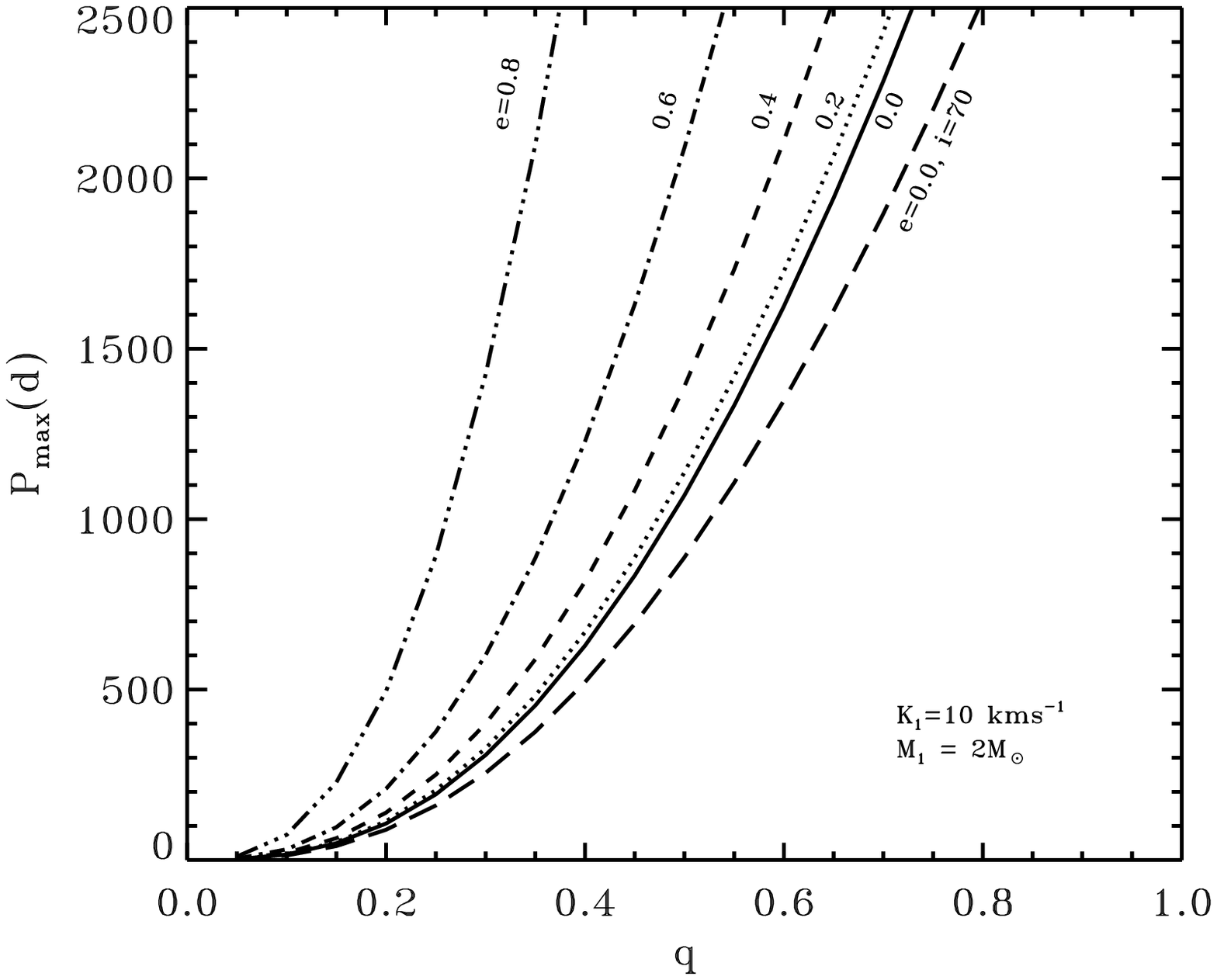}}
\caption{Left panel: the rotational velocities ($v \sin i$) of spectroscopic 
binaries with Am  components versus orbital period. The  hyperbolic
curves give the theoretical relation ($i=90^\circ$) for synchronized systems of 
$2\,M_\odot$ and different radii.
The discontinuous dotted line is the lower boundary of the ``avoidance zone", 
according to \cite{Bud96}, the vertical lines are the expected upper boundary 
for circularization, according to \cite{ZN03}: dotted, $R=2.1\,R_{\odot}$, $q=1$;  
dash-dotted, $R=3\,R_{\odot}$, $q=0.2$. 
Right panel: the longest orbital period of a SB1 binary, 
corresponding to the assumed minimum observable radial velocity amplitude 
$K_1=10$\,km\,s$^{-1}$, is plotted as function of mass ratio and eccentricity for 
primary mass $M=2\,M_{\odot}$ and inclination $i=90^\circ$}
\label{fig:vrot}
\end{figure}

 The recent survey of A-type star rotational velocities by \cite{Roal02} provides 
excellent material for the study  of  synchronization in binaries with Am components.
Figure~\ref{fig:vrot} shows the updated $v \sin i$ from R02 versus orbital period,
for the Am binaries listed in \cite{Bud96}, and the relations expected in case of 
spin-orbit synchronization.
The vertical lines are the periods  corresponding to the  fractional radius ($R/a$)   
that should  assure circularization (\cite{ZN03}) in the case of a 2\,$M_\odot$ MS-primary. 
The data are generally in good agreement with the expectations, with a marginal
indication of  radii larger than in normal stars (as the synchronization relations
should be upper envelopes, having assumed $i=90^\circ$).  
 
 The R02 velocities are generally higher than those measured by AM95 and used by Budaj; 
 the reason is a systematic effect in the velocity of standard stars used by AM95. 
 As a consequence the ``avoidance zone", found by \cite{Bud96}, for periods
$4<P<20$ days (i.e the region above the dotted line in Fig.~\ref{fig:vrot}) 
is no longer empty and it is not necessary an ad-hoc
mechanisms to explain it, such as the ``tidal mixing" introduced by the 
abovementioned author;  the decreasing  number of systems found in proximity 
of the synchronism lines can be due to a dependence on period of the braking efficiency.
 
 The other feature appearing in Figure~\ref{fig:vrot}, the lack of systems between
$180<P<800$ days, is quite probably a selection effect due to the low probability 
of discovery \textit{as a spectroscopic binary}. An idea of the relevance of selection 
effects can be derived from  the right panel of Fig.~\ref{fig:vrot}. This shows, 
for a $M=2\,M_{\odot}$ primary, the largest orbital period corresponding to a given 
minimum observable radial velocity amplitude, $K_1$. This is a function of mass 
ratio, $q$, and eccentricity, $e$  (in the hypothesis of inclination $i=90^\circ$). 
Even in the conservative hypothesis $K_1=10$ km\,s$^{-1}$ (but \cite{Hog92}, on the 
basis of a detailed study of selection 
effects for spectroscopic binaries,  suggests 25\,km\,s$^{-1}$ for A-type stars)
Figure~\ref{fig:vrot} indicates that the largest period of binaries detectable  by
spectroscopic surveys is of some hundred days (and many systems 
will escape detection, as  the distribution of mass ratio of single lined spectroscopic 
binaries is peaked around $q=0.2$). The gap in the period distribution could, therefore, 
be due to the transition between spectroscopic binaries, discovered as such, 
and visual binaries with known radial velocity.

In conclusion, a proper treatment of selection effects is a main issue to solve if
we want to extract correctly the rich information that A-type stars can provide on
binary secular evolution and that, in their turn, binaries can yield on A-type 
star properties.

\begin{acknowledgments}
The authors thank the IAU for grants covering the registration fees.
A.N and J.M. acknowledge ESA-PRODEX contract 15448/01/NL/SFe(IC)-C90135 and  IAP P5/36, 
and  C.M. a F-INAF program for funding. 
\end{acknowledgments}

\begin{discussion}
\discuss{}{Questions posponed to the panel discussion B (see pp. xxx - xxx).}

\end{discussion}
\end{document}